\documentclass{ceab}   

\usepackage{epsfig}     
\usepackage{graphicx}   
\usepackage{url}
\usepackage{natbib}     
\usepackage[T1]{fontenc} 

\usepackage{tikz}
\usepackage{xcolor}
\usepackage{xspace}

\newcommand{\pythia}{\textsc{Pythia}\xspace}
\newcommand{\angantyr}{\textsc{Angantyr}\xspace}
\newcommand{\corsika}{\textsc{Corsika}\xspace}
\usepackage{newunicodechar}
\newunicodechar{★}{\star}
\usepackage[english]{babel}
\usepackage{hyperref}
    
\def\runningtitle{Gaudu et al.: CORSIKA 8 with \pythia 8: Simulating Vertical Proton Showers} 
\def\articlenumber{7} 
\def\ppages{1--9} 

\begin{document}

\title{CORSIKA 8 with \pythia 8:\\ Simulating Vertical Proton Showers}

\author{C. Gaudu$^{1}$, M. Reininghaus$^{2}$ and F. Riehn$^{3,4}$\footnote{Now at \\
\textit{Technische Universität Dortmund, August-Schmidt-Straße 4, 44221 Dortmund, Germany\\
Ruhr-Universität Bochum, Universitätsstraße 150, 44801 Bochum, Germany}}\, for the \corsika 8 Collaboration\footnote{\scriptsize FAL: \href{https://gitlab.iap.kit.edu/AirShowerPhysics/corsika/-/wikis/Current-CORSIKA-8-author-list}{gitlab.iap.kit.edu/AirShowerPhysics/corsika/-/wikis/Current-CORSIKA-8-author-list}}
\vspace{2mm}\\
\it $^1$Bergische Universität Wuppertal,\\ 
\it Gaußstraße 20, 42119 Wuppertal, Germany\\
\it $^2$Karlsruhe Institute of Technology, Institute of Experimental Particle Physics,\\
\it Hermann-von-Helmholtz-Platz 1, 76344 Eggenstein-Leopoldshafen, Germany \\
\it $^3$Universidade de Santiago de Compostela,\\
\it Instituto Galego de Física de Altas Enerxías, \\
\it Rúa de Xoaquín Díaz de Rábago, E-15782 Santiago de Compostela, Spain \\
\it $^4$Laboratório de Instrumentação e Física Experimental de Partículas, \\
\it Avenida Professor Gama Pinto 2, 1649-003 Lisboa, Portugal \\
}

\maketitle

\begin{abstract}
The field of air shower physics, dedicated to understanding the development of cosmic-ray interactions with the Earth's atmosphere, faces a significant challenge regarding the muon content of air showers observed by the Pierre Auger Observatory, and numerous other observatories. Thorough comparisons between extensive air shower (EAS) measurements and simulations are imperative for determining the primary energy and mass of ultra-high energy cosmic rays. Current simulations employing state-of-the-art hadronic interaction models reveal a muon deficit compared to experimental measurements, commonly known as the ``Muon Puzzle''. The primary cause of this deficit lies in the uncertainties surrounding high-energy hadronic interactions.\\
In this contribution, we discuss the integration of a new hadronic interaction model, \pythia 8, into the effort to resolve the Muon Puzzle. While the \pythia 8 model is well-tailored in the context of Large Hadron Collider (LHC) experiments, its application in air shower studies remained limited until now. However, recent advancements, particularly in the \angantyr model of \pythia 8, offer promising enhancements in describing hadron-nucleus interactions, thereby motivating its potential application in air shower simulations.\\
We present results from EAS simulations conducted using \corsika 8, wherein \pythia is employed to model hadronic interactions.
\end{abstract}


\newpage
\section*{Introduction}

To gain insights into the ``Muon puzzle'' -- a persistent muon deficit observed in air shower simulations compared to measurements, e.g. from the Pierre Auger
Observatory -- several studies took place: from ad hoc modification of cross-section, multiplicity, and elasticity of hadronic interactions models (\cite{Ulrich:2010rg, Vicha:2023jup, Vicha:2022zvv, Ebr:2023nkf}), or by altering directly particle production, to performing a multi-parameter fit of model predictions against Auger data (\cite{Riehn:2024prp}). The purpose of this proceeding is to introduce another hadronic interaction model, \pythia 8 (\cite{Bierlich:2022pfr}), into the landscape of air shower simulations. Unlike the models most commonly used in this kind of study, namely EPOS-LHC, \textsc{Sibyll} 2.3d and \textsc{QGSJet}-II.04 (\cite{Pierog:2013ria, Riehn:2019jet, Ostapchenko:2010vb}), \pythia 8 was not built with the constraints of air showers physics in mind. 

In this work, we present the first results from the \emph{ongoing implementation} of \pythia 8 in \corsika 8 (\cite{Engel:2018akg, CORSIKA:2023jyz, Alameddine:2024cyd}) for vertical proton-induced showers. We discuss the longitudinal shower and energy loss profiles, as well as the lateral distributions and energy spectra of the particles at ground. 

\section{\pythia 8}

\pythia was conceived as a general purpose particle interaction model, optimized for describing LHC measurements, i.e. e$^{+}$e$^{-}$, pp, $\overline{\mathrm{p}}$p, pPb, PbPb interactions. The nuclear interactions are handled by the \angantyr model (\cite{Bierlich:2018xfw}, \cite{Bierlich:2021poz}) in \pythia 8, bridging the gap between high energy and heavy ion phenomenology. 

\begin{figure}[h!]
    \hspace{-0.8cm}
    \includegraphics[width=0.55\linewidth]{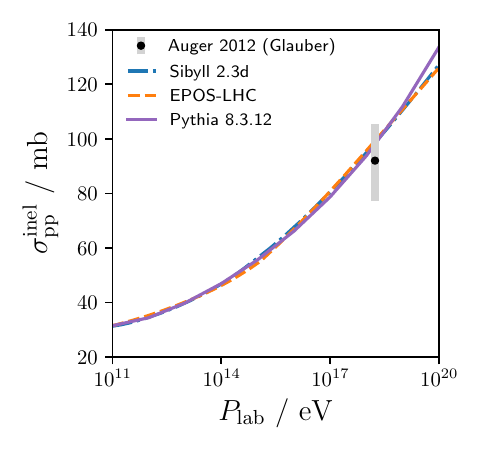}
    \hspace{-0.4cm} 
    \includegraphics[width=0.55\linewidth]{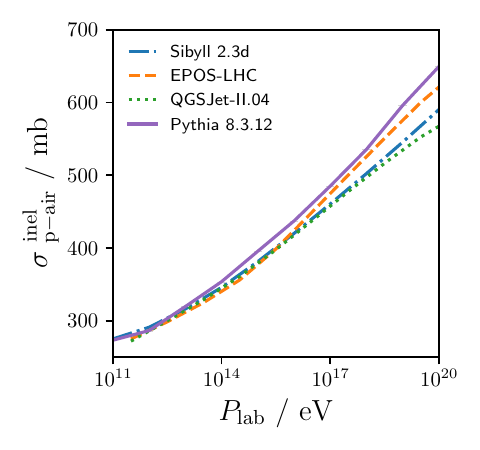}

    \vspace{-0.5cm} \caption{\small Lines: Inelastic cross-sections as a function of the momentum in the laboratory frame of the projectile for proton-proton (left) and proton-air (right) collisions, assuming the composition of air as 80\% $^{14}$N to 20\% $^{16}$O mix. Markers: inelastic cross-section for proton-proton collisions from~\cite{PierreAuger:2012egl}.}
    \label{fig:xsec}
\end{figure}

In Fig.\ref{fig:xsec}, the cross-sections for proton-proton and proton-air interactions, from $\sim$100 GeV to 100 EeV are displayed\footnote{Further collisions systems, such as $\pi$p and $\pi$-air, can be found in~\cite{Gaudu:2024mkp}.}. \pythia 8 is overall compatible with the commonly high energy models used in air shower physics: EPOS-LHC, \textsc{Sibyll} 2.3d and \textsc{QGSJet}-II.04. In the proton-proton case, all hadronic interaction models yield an extremely similar description, except for the highest energies where \pythia diverges, while in proton-air interactions, this divergence starts at the TeV scale already, establishing \pythia 8 as the model with the highest cross-sections. 

\section{\corsika 8 + \pythia 8}

The implementation of \pythia in \corsika 8 follows in the footsteps of previous studies by~\cite{Reininghaus:2022gnr} and~\cite{Reininghaus:2023ctx}, but now using the \angantyr model in \pythia 8.312 and its opportunities as discussed in~\cite{Sjostrand:2021dal} and~\cite{Gaudu:2024mkp}. A new feature allows for a single instance of \pythia to run, from which the beam system can be switched and the collision energy can vary on an event-by-event basis.

The plan is for all collisions that includes a nucleus to be treated by \angantyr\footnote{Using nuclear projectiles handled via \angantyr within C8 are a work-in-progress.}, while any other collisions are handled by \pythia. A comprehensive set of total and partial cross-sections for atmospheric collisions was tabulated to simulate interactions using \pythia within \corsika 8. This includes pions, kaons, protons, neutrons, other long-lived mesons and baryons as projectiles, as well as protons, $^{12}$C, $^{14}$N, $^{16}$O and $^{50}$Ar as targets.

For this work, 300 vertical air showers that were induced by protons with a primary energy of 10$^{17}$ eV were simulated using four different high-energy interaction models. The following energy thresholds for particle tracking were used: e$^\pm$/$\gamma$ particles at 10 MeV, hadrons and muons at 300 MeV. Electromagnetic particles with an energy below 10$^{-6}$ times the primary energy were combined statistically to reduce runtime; this thinning is detailed in~\cite{CORSIKA:2023jyz}. For technical reasons, the energy threshold to transition from the lower energy model FLUKA (\cite{Ballarini:2024uxz}) to \pythia 8 is currently set at 4 TeV, while the \corsika 8 default value is $\sim$80 GeV when using the other high energy models.

\subsection{Shower development}
The so-called longitudinal shower profile tracks the number of secondary particles as a function of atmospheric depth. It provides insights into the development of the shower as it propagates through the atmosphere.

\begin{figure}[h!] \hspace{-0.4cm}
    \includegraphics[trim={0 0 0 0.3cm}, clip,width=0.52\linewidth]{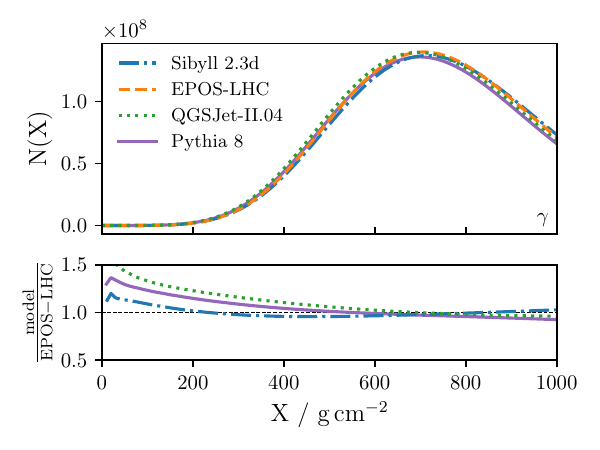}
    \hspace{-0.4cm}
    \includegraphics[trim={0 0 0 0.3cm}, clip,width=0.52\linewidth]{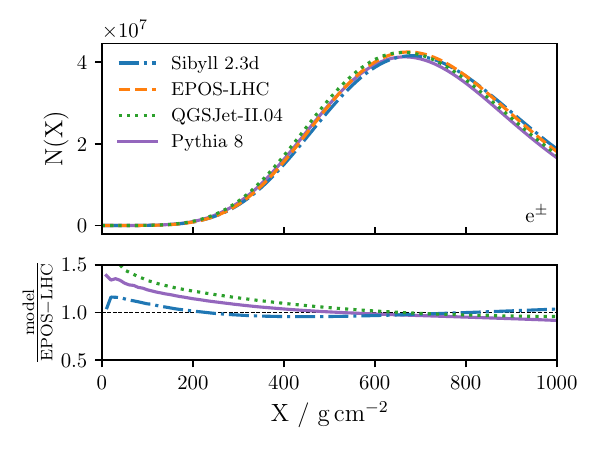} 

    \vspace{-0.2cm} \hspace{-0.4cm}
    \includegraphics[trim={0 0.1cm 0 0.3cm}, clip, width=0.52\linewidth]{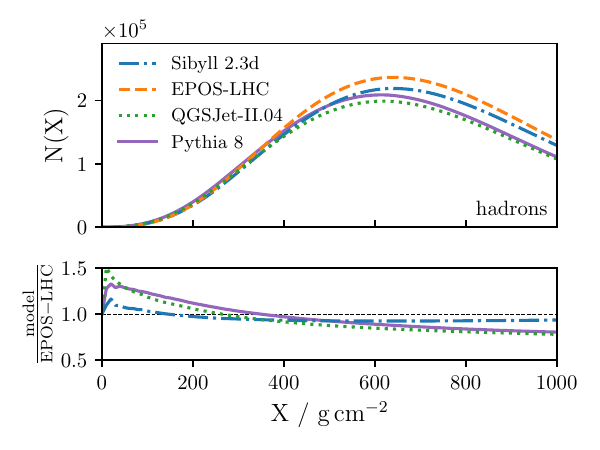}
    \hspace{-0.4cm}
    \includegraphics[trim={0 0.1cm 0 0.3cm}, clip, width=0.52\linewidth]{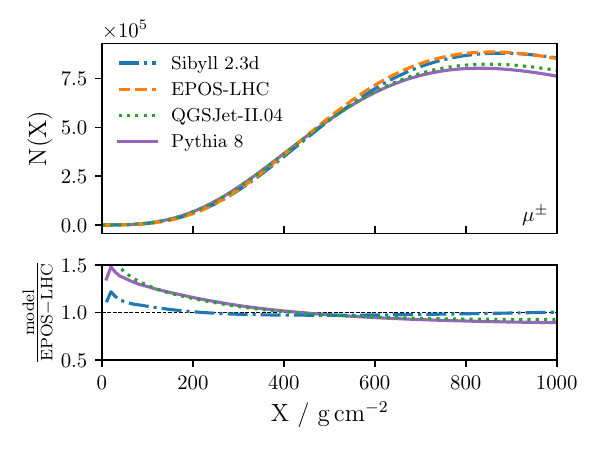}

    \vspace{-0.5cm} \caption{Average longitudinal shower profiles for vertical proton-induced 10$^{17}$ eV air showers for various particle types: photons (upper left), electrons/positrons (upper right), hadrons (lower left) and muons (lower right), using several high-energy hadronic interaction models and FLUKA as low-energy interaction model.}
    \label{fig:longitudinal_profile}
\end{figure}

The longitudinal profile of \pythia 8 showers for several particle types (photons, electrons/positrons, hadrons and muons) are displayed in Fig.\ \ref{fig:longitudinal_profile}. These profiles are compatible to those from~\cite{CORSIKA:2023jyz} and~\cite{Gaudu:2024kmh} when previously comparing \corsika 8 to \corsika 7 outputs, with interaction models EPOS-LHC, \textsc{Sibyll} 2.3d and \textsc{QGSJet}-II.04. 
All models display a \emph{crossing} behaviour relative to EPOS-LHC, with higher muon content predicted in the early stages of shower development, and similar or lower content at later stages. \pythia 8 and \textsc{QGSJet}-II.04 predict up to 50\% more muons at the early stages but end up with about 10\% fewer muons by $X \sim 1000$ g~cm$^{-2}$ compared to EPOS-LHC.

\begin{figure}[h!] \centering 
    \includegraphics[width=0.65\linewidth]{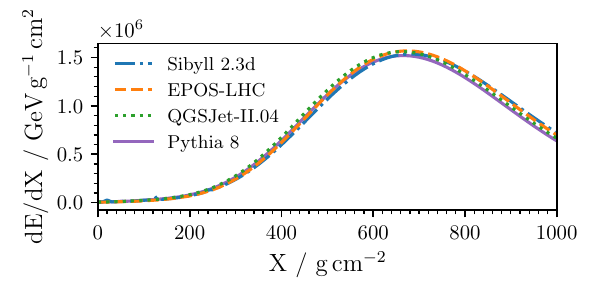} 
    \vspace{-0.6cm} \caption{Average total energy loss distributions for the same showers as in Fig.\ \ref{fig:longitudinal_profile}.} 
    \label{fig:dEdX}
\end{figure}

\begin{figure}[h!] \hspace{2.1cm}
    \includegraphics[width=0.55\linewidth]{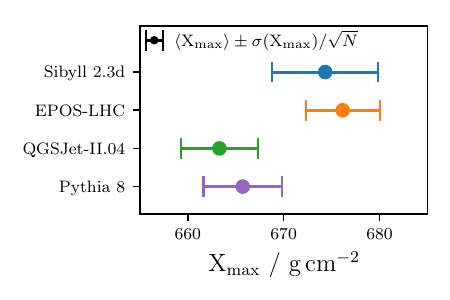}
    \vspace{-0.6cm} \caption{Average depth of the shower maximum for the same showers as in Fig.\ \ref{fig:longitudinal_profile}.}
    \label{fig:Xmax}
\end{figure}

In Fig.\ \ref{fig:dEdX} the total energy loss profile is shown, from which the depth of shower maximum, $X_\mathrm{max}$, is inferred and displayed in Fig.\ \ref{fig:Xmax}. The \pythia~8 prediction for the average depth of shower maximum is found to be slightly below the values for \textsc{Sibyll} 2.3d and EPOS-LHC, while being in good agreement with \textsc{QGSJet}-II.04.

It has been discussed in a recent study by~\cite{PierreAuger:2024neu} that a deeper $X_\mathrm{max}$ would lead to a larger number of muons at ground, making the description of \pythia 8 and \textsc{QGSJet}-II.04 the least attractive toward a resolution of the Muon Puzzle in this scenario. 

\subsection{Particles at ground}

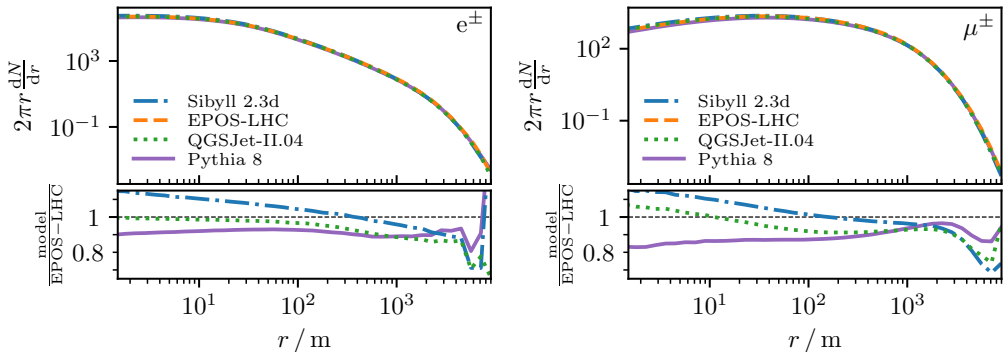
\begin{figure}[h!] \hspace{-0.8cm}
    \includegraphics[trim={0 0.1cm 0 0}, clip, width=0.55\linewidth]{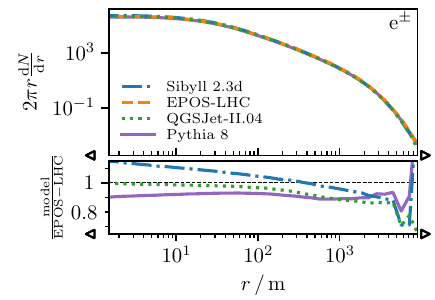}
    \begin{tikzpicture}[remember picture, overlay]
        \node[fill=white, minimum width=0.2cm, minimum height=0.2cm] at (-0.19, 0.98) {};
        \node[fill=white, minimum width=0.2cm, minimum height=0.2cm] at (-0.19, 2.2) {};
        \node[fill=white, minimum width=0.2cm, minimum height=0.2cm] at (-5.5, 0.98) {};
        \node[fill=white, minimum width=0.2cm, minimum height=0.2cm] at (-5.5, 2.2) {};
    \end{tikzpicture} \hspace{-0.45cm}
    \includegraphics[trim={0 0.1cm 0 0}, clip, width=0.55\linewidth]{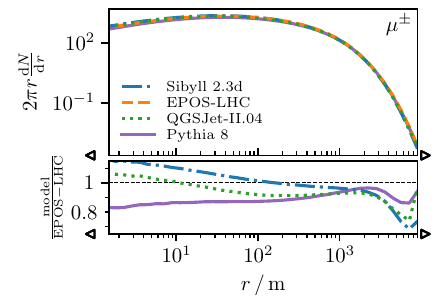} 
    \begin{tikzpicture}[remember picture, overlay]
        \node[fill=white, minimum width=0.2cm, minimum height=0.2cm] at (12.74, 1.4) {};
        \node[fill=white, minimum width=0.2cm, minimum height=0.2cm] at (12.74, 2.7) {};
        \node[fill=white, minimum width=0.2cm, minimum height=0.2cm] at (7.4, 1.4) {};
        \node[fill=white, minimum width=0.2cm, minimum height=0.2cm] at (7.4, 2.7) {};
    \end{tikzpicture}
    \vspace{-0.6cm} \caption{Average lateral distributions of electrons/positrons (left) and muons (right) at sea level for the same showers as displayed in Fig.\ \ref{fig:longitudinal_profile}.}
    \label{fig:ldf}
\end{figure}

The lateral distribution function shows the particle number density as a function of distance from the shower impact point at a specific altitude. Unlike the longitudinal profile, it represents the shower at a particular stage of its development. As shown in Fig.\ \ref{fig:ldf}, \pythia 8 predicts fewer electrons, positrons, and muons from near the shower core up to $\sim 1$~km, at sea level, in comparison to the other hadronic interaction models.

\begin{figure}[h!]
    \hspace{-0.8cm}
    \includegraphics[width=0.55\linewidth]{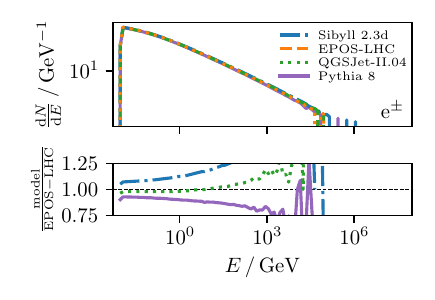}
    \hspace{-0.45cm}
    \includegraphics[width=0.55\linewidth]{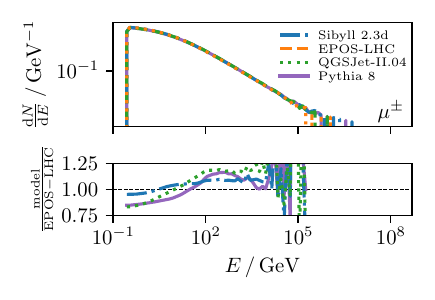}

    \vspace{-0.6cm} \caption{Average energy spectra of electrons/positrons (left) and muons (right) at sea level for the same showers as displayed in Fig.\ \ref{fig:longitudinal_profile}. High-energy interactions are modeled by \pythia 8 above 4 TeV and by other high-energy models above 80 GeV, with FLUKA handling interactions below these thresholds.}
    \label{fig:espect}
\end{figure}

Another observable related to the secondaries reaching the ground is the average energy spectrum of charged leptons depicted in Fig.\ \ref{fig:espect}. It seems that \pythia produces fewer electrons and positrons reaching the ground compared to the other interaction models. On the other hand, it offers a similar description for secondary muons at the ground as \textsc{QGSJet}-II.04, up to the TeV scale, with a slight offset towards higher energy muons. Above $\sim 10$ TeV, fluctuations from the limited size of the set of simulated showers are dominant for all models.

\newpage
\section*{Conclusion}

In summary, this work shows that full air shower simulations can now be performed using \pythia 8 within the \corsika 8 framework. Focusing on vertical proton-induced showers at 10$^{17}$ eV, we found that \pythia 8 predicts shower observables that are in agreement, for the most part, with those from other hadronic interaction models. The longitudinal profiles are generally consistent across the models. While \pythia and EPOS-LHC exhibit some differences in the depth of shower maximum, \pythia remains compatible with both \textsc{Sibyll} 2.3d and \textsc{QGSJet}-II.04. The lateral distributions and energy spectra at the ground exhibit distinct behaviours but remain broadly comparable. This highlights the potential of \pythia 8 as a viable tool for air shower simulations.

The strengh of \pythia resides in its tuning opportunities, allowing in a friendly and well-documented manner its users to modify internal parameters of the model with a strong impact on air shower observables. Therefore, once the investigations of vertical showers with nuclear primaries and inclined air showers for both proton and nuclear primaries are finalized, a tuning study of \pythia will be the next logical step towards refining \pythia in relation to the Muon Puzzle. A forward physics tune of \pythia was previously presented by~\cite{Fieg:2023kld}, while ongoing efforts for a tune using fixed-target measurement are discussed by~\cite{Gaudu:2024mkp}. 

\section*{Acknowledgements} 
C. Gaudu acknowledges funding from the German Research Foundation (DFG, Deutsche Forschungsgemeinschaft) via the Collaborative Research Center SFB1491: Cosmic Interacting Matters – from Source to Signal (F4) -- project no. 445052434. F. Riehn received funding from the European Union’s Horizon 2020 research and innovation programme under the Marie Skłodowska-Curie grant agreement No. 101065027. We thank T. Sjöstrand and the \pythia 8 collaborators for their support in the implementation of \pythia 8 in \corsika 8.

\bibliographystyle{apalike}
\bibliography{pythia_8}

\end{document}